\documentclass[pra,aps,twocolumn,showpacs,superscriptaddress,floatfix,amsmath,amssymb,nofootinbib]{revtex4}
\usepackage{amsfonts}
\usepackage{graphicx} 
\newcommand{\beq}{\begin{equation}}
\newcommand{\eeq}{\end{equation}}

\newcommand{\beqa}{\begin{eqnarray}}
\newcommand{\eeqa}{\end{eqnarray}}
\def\ra{\rangle}
\def\la{\langle}
\begin{document} 
\title{Enhanced observability of quantum post-exponential decay using distant detectors}
\author{E. Torrontegui}
\email{jg.muga@ehu.es}
\affiliation{Departamento de Qu\'{\i}mica-F\'{\i}sica, Universidad del Pa\'{\i}s Vasco, 
Apdo. 644, Bilbao, Spain}
\author{J. G. Muga}
\email{jg.muga@ehu.es}
\affiliation{Departamento de Qu\'{\i}mica-F\'{\i}sica, Universidad del Pa\'{\i}s Vasco, 
Apdo. 644, Bilbao, Spain}
\author{J. Martorell}
\email{martorell@ecm.ub.es}
\affiliation{Departament d'Estructura i Constituents de la Materia, 
Facultat F\'{\i}sica, \\ 
University of Barcelona, Barcelona 08028, Spain}
\author{D. W. L. Sprung}
\affiliation{Department of Physics and Astronomy, McMaster University, \\
Hamilton, Ontario L8S 4M1 Canada}


\begin {abstract} 
We study the 
elusive transition from exponential to post-exponential (algebraic) decay 
of the probability density of a quantum particle emitted by  
an exponentially decaying source, in one dimension. The main finding is that 
the probability density at the transition time, and thus its observability, 
increases with the distance of the detector from the source, up to a critical distance 
beyond which exponential decay is no longer observed.  
Solvable models provide explicit expressions 
for the dependence of the transition 
on resonance and observational parameters,
facilitating the choice of optimal conditions.    
\end{abstract}  	
\maketitle

\section{Introduction}
Exponential decay results when the  
decay rate depends linearly on the surviving population. It is 
observed in many quantum systems, in virtually all fields of physics 
(particle, nuclear, atomic, molecular, and condensed matter), but its 
microscopic derivation from the Schr\"odinger equation is not as straighforward  
as in classical physics, due to initial state reconstruction \cite{isr}, 
and deviations are predicted at both short and long times \cite{FGR78,FP08,book}.
The \index{quantum Zeno effect} Zeno effect,   
in particular, is associated with the short time deviation. 
The deviations at long times are 
less often  discussed, and constitute the central topic of this paper.

On the experimental side, the post-exponential regime,
which is typically algebraic, has been elusive. There have been 
many attempts to produce evidence of post-exponential decay, with little success 
\cite{Nik68,No88,Gr88,No95,Nghiep98}. 
It has been argued that repetitive measurements on the same system, 
or simply the interaction with the environment would lead to 
persistence of the exponential regime to times well beyond those expected in an isolated system 
\cite{FGR78,Greenland86,Law02b}. Nevertheless a recent measurement of transitions 
from  exponential to post-exponential decay of excited organic molecules in solution \cite{RHM06} 
appears to contradict this expectation, and has triggered renewed interest in the topic.  

Aside from the non-trivial challenge of understanding and extending these 
experimental results, there are many reasons for studying post-exponential decay: 
Winter, for example, argued that hidden-variable 
theories could produce observable effects in samples that have decayed for 
many life-times \cite{Winterh};  more recently Krauss and Dent have  
pointed out that late-time decay may have important cosmological implications \cite{KD08};  
decay at long times is quite sensitive to delicate measurement and/or environmental effects, 
so it is a testing ground for theories of these processes;  
at a  fundamental level, the form of long time deviations from exponential decay might 
distinguish between standard, Hermitian quantum mechanics \cite{DS84}, 
and modifications with a built-in microscopic arrow of time \cite{NB77,Nicolaides02}; 
from a practical perspective, Norman pointed out that post-exponential decay could set a 
limit to the validity of radioactive dating methods \cite{No95};   
we have also argued that the deviation could be used to characterize certain cold atom traps \cite{MMS08}. 
Indeed, due to technological advances in lasers, 
semiconductors, nanoscience, and cold atoms, microscopic 
interactions are now relatively easy to manipulate: this makes decay 
parameters controllable, and post-exponential decay more accessible to 
experimental scrutiny and/or applications. For example, under 
appropriate conditions it could become the dominant regime 
and be used to speed-up decay, implementing an Anti-Zeno effect \cite{Lewenstein}.   
In addition, recent experiments on periodic 
waveguide arrays provide a classical, electric field analog of a quantum 
system with exponential decay \cite{Lon06a,VLL07}. These experiments 
may also reach the post-exponential region in a particularly simple way. 

Most of the proposals to enhance post-exponential decay, 
making it more visible by advancing its onset time, are based on the 
idea of observing the survival probability for a resonance close to threshold 
(``small Q-value decay'') \cite{Jittoh,GV06,RLE82,ZRL84,Lewenstein,RHM06}, or more precisely, 
with energy release comparable to resonance width. This is 
a rare configuration in 
naturally decaying systems, such as radioactive isotopes \cite{RHM06},
or it leads to difficult measurements \cite{ZRL84,Greenland86,Kelkar}, 
and realizing it in artificial structures, although possible in principle,
is still a pending task \cite{Jittoh}. The only realization so far 
may be the experiment on organic molecules in solution, mentioned above \cite{RHM06},
but this is a rather complex system and its analysis from first principles has not yet 
been performed, so the true mechanism behind the observed data 
remains to be confirmed.   Another idea is to use the escape of interacting cold atoms 
from a trap, in the strongly interacting Tonks-Girardeau regime, to 
enhance the signal \cite{DDGMR06}. Again, the corresponding experiment
has not yet been done. 
We also mention the proposal of Kelkar and coworkers \cite{Kelkar}
to deduce, rather than observe directly,  the long-time decay characteristics 
from phase-shift data.   

\begin{figure}[ht]
\begin{center}
\includegraphics[height=5cm,angle=0]{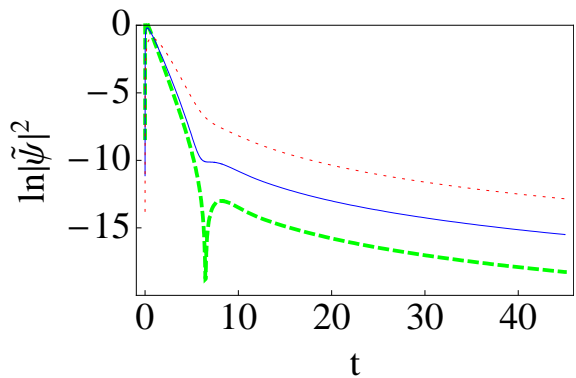}
\end{center}
\caption{\label{exact}
(Color online) Probability density versus $t$ for a fixed value of $k_{0}=1-0.5i$. Three 
detector positions $x$ have been considered:  $x=1.5$ (red dotted line), $x=0.4$ (blue solid line), and $x=0.1$ (green dashed line). 
The dimensionless variables and the model are explained in Sec. II.}
\end{figure}

In the present paper we propose a quite different strategy with the 
important advantage of simple implementation. 
We show, using solvable decay models, that by increasing the distance of the observation point 
from the source, the transition from exponential to  
post-exponential decay occurs at higher probability densities, and thus becomes 
more easily observable. Figure \ref{exact} shows the probability density  
versus time, for three positions, and illustrates the density increase at the transition 
as the observation position $x$ is increased. (The details of the  model will be explained in Sec. II.) 
The observability of the transition may additionally be enhanced by amplifying the signal with a 
macroscopic Bose-Einstein condensate wavefunction as explained in the final discussion. The combination 
of enhancing factors makes it possible to predict the observability of the transition for longer lifetimes 
than those implied by the above mentioned ``small Q-value'' condition.     

Finally, note that when the distance is too large the  exponential decay regime  
disappears, and with it the transition, so that only non-exponential decay should be observed \cite{Newton}.     

In Section II a simple decay model is introduced and solved exactly. Section III
provides approximate expressions useful to define the transition 
from exponential to post-exponential decay discussed in Section IV. 
The final section provides examples of possible 
physical realizations and compares the results with those of a discretized site model. 
\section{Exponentially decaying source model and exact solution}
There is a long and fruitful tradition of analytically solvable models for studying 
diverse aspects of single particle quantum dynamics. 
Closest to the model proposed here are several ``source models'' to examine 
quantum transients (propagation of signals, diffraction in time, forerunners) and tunneling dynamics of evanescent waves \cite{MB}. 
Many works have been carried out within this approximation to study neutron interferometry \cite{FMGG90,BSCK92},
atom-wave diffraction \cite{BZ97}, tunneling dynamics \cite{Stevens,Ranfa90,Ranfa91,Mor92,BT98,MB,DMRGV},
absorbing media \cite{DMR04}, different aperture functions \cite{FMGG90,BZ97,DM05}, 
and to model the dynamics of a guided atom laser \cite{DMM07}.

In these ``source models'', unlike the usual initial value problem, the wave function is specified for  
all times at one point of coordinate space. (The connection between initial value problems and source 
boundary conditions has been discussed in \cite{BEM01}). A simple, solvable 1D case 
corresponds to switching-on the source suddenly at $T=0$, with a well defined ``carrier frequency'' $\Omega_0>0$,  
\beq					
\Psi(0,T)=\Theta(T)e^{-i\Omega_0 T},
\eeq
zero external potential, $V=0$, and a vanishing wave everywhere (for all $X$) before 
the source is switched on, i.e. for $T<0$. 
The solution of the Schr\"odinger equation for $X>0$ is identical to that for the ``Moshinsky shutter'',
i.e. a wave created by the sudden opening of an opaque shutter which totally reflects
an incident plane wave with reflection amplitude $r=1$ \cite{BZ97} prior to $T=0$.    
This solution shows the ``diffraction in time'' phenomenon of 
a fairly  well-defined wavefront advancing with velocity $K_0\hbar/m$, where
\beq					
K_0=\sqrt{2m\Omega_0/\hbar}
\eeq
and $m$ is the mass of the particle. 
Negative values of $\Omega_0$ correspond to  particle injection with energy 
below the potential threshold and 
may also be considered: interesting transients are produced in this way and, at long times,   
a stationary evanescent wave which penetrates into the classically forbidden 
region \cite{BT98,MB}.    
Here we shall work out a case that has not yet been examined, namely, 
a complex carrier frequency $\Omega_0=\Omega_{0R}+i\Omega_{0I}$. 

We assume a negative imaginary part, $\Omega_{0I}<0$, 
so that the density at the source point decreases exponentially,  
\beq
|\Psi(0,T)|^2=e^{-2|\Omega_{0I}| T}.  
\eeq
The real part $\Omega_{0R}$ is chosen to be positive so as to
generate travelling waves released from the source, and thus  mimic 
the behaviour of resonance decay above threshold, in the outer region
(outside the interaction region that holds the 
resonance).\footnote{There are no resonances below the energy threshold, so we disregard 
the case $\Omega_{0R}<0$, even though it could be treated formally
in parallel with the $\Omega_{0R}>0$ case. 
The evanescent case $\Omega_{0R}<0$ might represent approximately the decay
of a resonance in a tunnelling region of space far from the
particle release to the outer region. The main differences with respect to 
the travelling case will be pointed out in a later footnote.}  
  
This provides a simple (minimal) solvable model 
for an exponentially decaying system with a long lived resonance, 
while avoiding a detailed description of the interaction region
where the decaying system is prepared. For many applications,
and in particular during the exponential decay of the initial state,  
these details are unnecessary in the outer region, 
which in our case is represented by the positive half-line. 
We have also checked, see the final discussion,  the broad validity of the results obtained  
using a different, complementary, norm-conserving, discretized lattice model. In it, the particle 
is prepared in the first site, which plays the role of the initial trap from which it escapes.    
This discretized model can be realized in periodic waveguide arrays, 
making use of an electric field analog of the quantum system \cite{Lon06a,VLL07}. 
\subsection{The dimensionless Schr\"odinger equation}
The dynamics of a particle in free space, $V=0$,
is given by the Schr\"odinger equation
\beq					
\label{eq1}
i\hbar\frac{\partial\Psi}{\partial T}=-\frac{\hbar^2}{2m}\frac{\partial^2\Psi}{\partial X^2}.
\eeq
Let us first reduce the number of variables by introducing
dimensionless quantities for position, time and wave function,
in terms of some characteristic length $L$: 
\beqa					
\label{eq2}
x&=&X/L, 
\\
\label{eq3}				
t&=&\frac{T\hbar}{2mL^2}, 
\\
\label{eq4}				
\psi(x,t)&=&\sqrt{L}\,\,\Psi(X,T).
\eeqa
$L$ may be chosen for convenience and we take it as   
$K_{0R}^{-1}$, where $K_0=(2m\Omega_0/\hbar)^{1/2}=K_{0R}+iK_{0I}$ is the dimensioned complex wave number and $K_{0R}$ its real part. 
The branch cut is 
drawn just below the negative real axis so that for $\Omega_{0I}<0$ the imaginary part 
of $K_{0I}$ is negative, $K_{0I}:={\rm{Im}}(K_0)<0$.  
Defining dimensionless wavenumber and carrier frequency as $k_0=K_0L$, and 
$\omega_0=\Omega_0 2mL^2/\hbar=k_0^2$, 
this election fixes the real part of the dimensionless wave number to be unity, $k_{0R}=1$, and the dimensionless 
dispersion equation becomes 
\beq					
\label{eq5c}
\omega_{0}=k_{0}^{2}=(1+ik_{0I})^2,   
\eeq
where $-1<k_{0I}<0$. 
In terms of these new variables, the Schr\"odinger equation (\ref{eq1}), is now 
\beq					
\label{eq5}
i\frac{\partial\psi(x,t)}{\partial t}=-\frac{\partial^2\psi(x,t)}{\partial x^2}, 
\eeq
and the dimensionless continuity equation is
\beq
\label{cont}
\frac{\partial \rho(x,t)}{\partial t}+\frac{\partial J(x,t)}{\partial x}=0,
\eeq
where 
\beqa					
\rho(x,t)&=&|\psi(x,t)|^2,
\\					
J(x,t)&=&2\,{\rm Im}\left[\psi^{*}(x,t)\frac{\partial\psi(x,t)}{\partial x}\right],
\eeqa
are the dimensionless probability and current densities.
\subsection{Explicit solution\label{obtent}}
We are interested in waves that decay exponentially in time so the Schr\"odinger equation must  
satisfy  boundary conditions at the origin 
\beq					
\label{eq6}
\psi(0,t)=e^{-i\omega_0 t}\theta(t),\;\;\;\;\omega_{0R}>0,\, \omega_{0I}<0, 
\eeq
and at infinity, where the wave remains bounded as $x\to\infty$.  
which guarantees that the wave function does not diverge at $x=\infty$.

To find the solution for $x>0$ we make the ansatz
\beq					
\label{psixt}
\psi(x,t)=\int_{-\infty}^\infty\!\!d\omega\, A(\omega)e^{i\sqrt\omega x}e^{-i\omega t},
\eeq
and determine the function $A(\omega)$ from the boundary condition (\ref{eq6}), 
\beq					
\psi(0,t)=e^{-i\omega_0 t}\theta(t)=\int_{-\infty}^\infty\!\!d\omega\, A(\omega)e^{-i\omega t}. 
\eeq
Inverting the Fourier transform we obtain 
\beq						
A(\omega)=\frac{i}{2\pi(\omega-\omega_0)},
\eeq
The resulting integral for $\psi(x,t)$ is done most easily in the complex $k$ plane,    
\beq					
\label{eq15}
k = \sqrt{\omega}.  
\eeq
We may rewrite Eq. (\ref{psixt}) for the wave function as
\beq					
\label{eq16}
\psi(x,t)=\frac{-1}{2\pi i}\int_{\downarrow_{\rightarrow}}\!\!dk\, e^{-i(k^2t-kx)}\left(\frac{1}{k+k_0}+\frac{1}{k-k_0}\right),
\eeq
where the branch cut for $k=\sqrt{\omega}$ (or $k_0=\sqrt{\omega_0}$)  is chosen as before, 
slightly below the negative real axis. The integration path runs first from $i\infty$ to $\infty$ 
in the complex $k$-plane, but using Cauchy's theorem it can be deformed into the diagonal path 
$\Gamma_+$ in the second and fourth quadrants 
passing above the pole at $-k_0$ in the second quadrant,
\beq					
\label{eq18}
\psi(x,t)=\frac{-1}{2\pi i}\int_{\Gamma_+}\!\!dk\, e^{-i(k^2t-kx)}\left(\frac{1}{k+k_0}+\frac{1}{k-k_0}\right), 
\eeq
which may be split into two integrals of the form 
\beq					
\label{eq19}
\mathfrak{I}=\int_{\Gamma_+}\!\!dk\,\frac{e^{-i(ak^2+bk)}}{k-k_p},\;\;\; a>0,
\eeq
The saddle point of the exponent is at $(k_R=-b/2a,k_I=0)$ and the steepest descent path is the straight line $k_R=-(k_I+b/2a)$. 
These integrals have been studied in \cite{MB} by contour deformation into the steepest descent path. 
Independently of the pole position at $u_p=u(k_p)$, the result is 
\beq					
\label{eq27}
\mathfrak{I}=-i\pi e^{ib^2/4a}w(-u_p), 
\eeq 
where $w(z):=e^{-z^{2}}{\rm{erfc}}(-i z)$ is the Faddeyeva- (or ``$w$")-function \cite{FT61,AS65}, and $u_p=u(k_p)$, where  
\beq					
\label{eq21}
u(k)=\sqrt{\frac{a}{2}}(1+i)(k+b/2a),  
\eeq
which becomes real along the (straight) steepest descent line, and vanishes at the saddle point. 
  
Now we return to Eq. (\ref{eq18}) where $a=t$ and $b=-x$, so the saddle point is at $k_s=x/(2t)$: 
this corresponds to the velocity of a classical particle that arrives at $x$ at time $t$ 
having departed from the origin at $t=0$. The steepest descent path is the straight line 
$k_I=-[k_R-x/(2t)]$. Applying Eq. (\ref{eq27}), the wave function can finally be written as
\beq					
\label{eq28}
\psi(x,t)=\frac{1}{2}e^{ik_s^2 t}\left[w(-u_0^{(+)})+w(-u_0^{(-)})\right],
\eeq
where
\begin{eqnarray}			
u_0^{(+)}&=&u(k_0)=(1+i)\sqrt{\frac{t}{2}}k_0(1-\tau/t),
\label{eq29} 
\\  					
u_0^{(-)}&=&u(-k_0)=-(1+i)\sqrt{\frac{t}{2}}k_0(1+\tau/t),
\label{eq30}
\end{eqnarray}
and the modulus of the complex time  
\beq					
\label{eq31}
\tau=\frac{x}{2k_0}
\eeq
generalizes the B\"uttiker-Landauer traversal time, at least formally, 
in the present context \cite{MB}.   
\subsection{Wave function normalization\label{normali}}
For a fair comparison between different $k_{0}$ values, the total number of emitted particles must be kept 
fixed for all $k_0$, so the wave fuction must be normalized. The normalization constant can be 
obtained from the continuity equation  (\ref{cont}) by integration, 
\beq					
N_{+}(t=\infty)=\int_{0}^{\infty}\!\!dt\,J(0,t),
\eeq
where 
\beq					
N_+(t)=\int_{0}^{\infty}\!\!dx\,\rho(x,t)
\eeq
is the norm, counting particles emitted to $x \geq 0$.

As we assume that one particle is emitted as $t\rightarrow\infty$, the normalized
%
%
wave function, denoted by a tilde, is  
\beq					
\tilde \psi(x,t)=\frac{1}{\sqrt{\int_{0}^{\infty}\!\!dt\,J(0,t)}}\psi(x,t).
\eeq
\section{Asymptotic behaviour}
\begin{figure}[ht]
 \begin{center}
\includegraphics[height=5cm,angle=0]{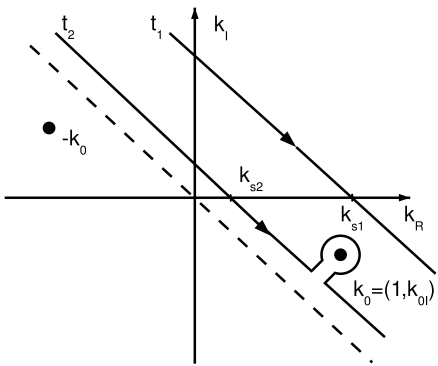}
 \end{center}
\caption{\label{kplane}
$k$-plane with contour deformations along the steepest descent paths
(SDP) (the straight lines) at times $t_2>t_1$, and saddle points $k_{s1}$, $k_{s2}$.  
At $t_1$, the $k_0$ pole has not yet been crossed, while at $t_2$ the pole has been crossed 
and encircled by the keyhole contour. The dashed line is the SDP for $t\sim\infty$, showing  
that the pole at $-k_0$ is never crossed.} 
\end{figure}

The exact solution is useful but not necessarily physically illuminating, whereas approximations 
often provide a more transparent physical interpretation. 
To this end we separate each $w$-function term into the contribution of 
the saddle point $k=k_s$,  and a pole contribution. 
The dominant contribution of the saddle is obtained from Eq. (\ref{eq18}) 
by setting $k=k_s$ in the denominators and integrating along the steepest descent path, 
\beqa				
\label{eq32}
\psi_s(x,t) &=& \frac{1}{2i\sqrt{\pi}}\left(\frac{1}{u^{(+)}_{0}}+\frac{1}{u^{(-)}_{0}}\right)
\nonumber \\ 
&=& \sqrt{\frac{2t}{\pi}}\frac{\tau e^{ik_s^2t}}{(i-1)k_0(t^2-\tau^2)}.
\eeqa
After the steepest descent path crosses the pole $k_0$ when ${\rm Im}{u_0^{(+)}}=0$, 
its residue $\psi_0(x,t)$ must be added to the saddle wave function $\psi_s$, 
see Fig. \ref{kplane},  
\beq				
\label{eq33}
\psi_{0}=e^{ik_s^2t}e^{-u_{0}^{(+)2}}=e^{-i\omega_0 t}e^{ik_0x}.
\eeq
(Note that the pole at $-k_0$, in the fourth octant, is never crossed.) 
A travelling wave component moving rightwards is generated. For fixed $t$ the exponential decay is imprinted in $e^{ik_0x}$: 
because of the negative imaginary part of $k_0$ the wave increases  
rightwards in coordinate space until the $x$ value satisfying ${\rm Im}{u_0^{(+)}}=0$.
Of course the full wave does not show this sharp edge but a smoothed one. 

It follows that the total wave function can be approximated as 
\beq					
\label{eq36}
\psi(x,t)=\psi_s(x,t)+\psi_{0}(x,t)\,\Theta[\mbox{Im}(u_{0}^{(+)})].
\eeq
To see the conditions for validity of this expression, 
note that the first and the exponential terms in the asymptotic series expansion for  $w(z)$ at large $|z|$,   
\beqa					
\label{eq38}
w(z)&\sim&\frac{i}{\sqrt{\pi}z}\left[1+\sum_{m=1}^{\infty}\frac{1\times3\times\ldots\times(2m-1)}{(2z^2)^m}\right]
\nonumber\\
&+&2e^{-z^2}\,\,\Theta[-{\rm Im} (z)], 
\eeqa
do reproduce Eq. (\ref{eq36}). 
In our case, large $z$ means large $u_0^{(\pm)}$. 
Their moduli can be written  as 
\beq					
|u_0^{(\pm)}|=\sqrt{x|k_0|}\sqrt{\frac{t}{2|\tau|}\pm\frac{1}{|k_0|}
+\frac{|\tau|}{2t}},
\eeq
which take large values in certain 
conditions, in particular at times short and long compared to $|\tau|$.

\section{Transition from exponential to post-exponential decay}
The criterion which sets the time scale of the transition,  
is that the moduli of pole and saddle contributions to the total density are equal. 
We thus identify the transition as the space-time point where their ratio $R(x,t)$ is one,  
\begin{eqnarray}				
|\psi_{0}|^2 &=& e^{2 {\rm{Im}}(\omega_0 t-k_0 x)},
\label{eq41}
\\						
|\psi_s|^2 &=& \frac{t |\tau|^2}{\pi|k_0|^2 [t^4+|\tau|^4-2t^2{\rm Re}(\tau^2)]},
\label{eq42}
\\						
R(x,t)&=&\frac{|\psi_{0}|}{|\psi_s|}=\frac{2\sqrt{\pi}|k_0|^2\,t^{3/2}}{x}e^{{\rm Im}(\omega_0t - k_0x)}
\nonumber\\
&\times&\sqrt{1+\frac{|\tau|^4}{t^4}-2{\rm Re}\left(\frac{\tau^2}{t^2}\right)}. \label{eq43}
\end{eqnarray}
In addition, the condition 
${\rm Im}(u_{0}^{(+)})>0$ (the pole has been crossed by the steepest descent path) must be verified.

Solving $R(x,t)=1$ with the assumption $t>>|\tau|$, we obtain an expression for 
the transition time $t_p$ for fixed $x$,
\beq						
\label{eq45}		
t_p^{3/2}=\frac{xe^{-{\rm Im}(\omega_{0} t_{p}- k_{0}x)}}{2\sqrt{\pi}|k_0|^2}.
\eeq
The transition is generally not sharply defined in the exact density, and is characterized 
by interference fringes due to the superposition of saddle and resonance contributions.   

\begin{figure}[ht]
 \begin{center}
  \includegraphics[height=6cm,angle=0]{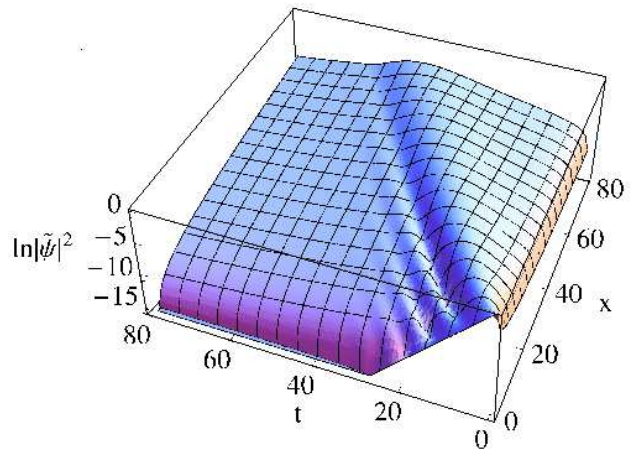}
 \end{center}
\caption{\label{3dim}
Overview of the Log probability density in the $t$, $x$ plane, for fixed $k_0=1-0.15i$.}
\end{figure}

\begin{figure}[ht] 
 \begin{center}
  \includegraphics[height=5cm,angle=0]{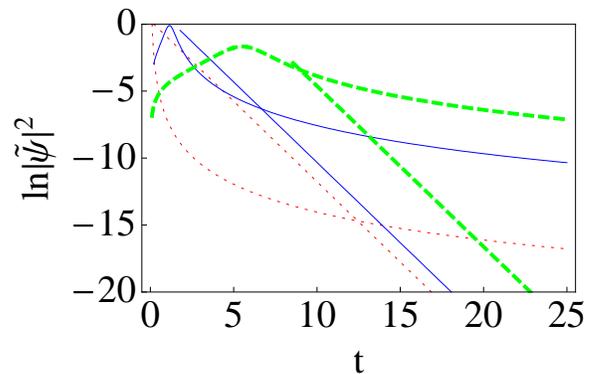}
 \end{center}
\caption{\label{denx}
(Color online) Probability density versus time, for three $x$ values, at fixed $k_0 = 1 -0.3 i$.  
The crossing point between the pole (straight line) and saddle terms defines the transition time $t_p$.
Dotted (red) line: $x=0.1$, crossing at $t_p = 12$; 
solid (blue) line: $x=2.5$, crossing at $t_p = 7$:  and 
dashed (green) line: $x=12$, crossing at $t_p = 9$.}
\end{figure}

In the following we shall investigate the behaviour of the probability density at the transition point $t_p$,  
when the parameters $x$ and/or the imaginary part $k_{0I}$ of $k_0$ are varied. 
First we fix $k_0$ and observe the density as a function of $t$ for different $x$-values. 
A global 3D plot is shown in Fig. \ref{3dim},  
in which we chose $k_{0I}$ to highlight the dominance of exponential decay at small $x$ and post-exponential at large $x$.
The pole (straight lines) and saddle densities for three values of $x$ 
are shown in Fig. \ref{denx}, in which $x$ increases from the dotted to solid to dashed lines. 

%
%
\begin{figure}[ht]
 \begin{center}
  \includegraphics[height=5cm,angle=0]{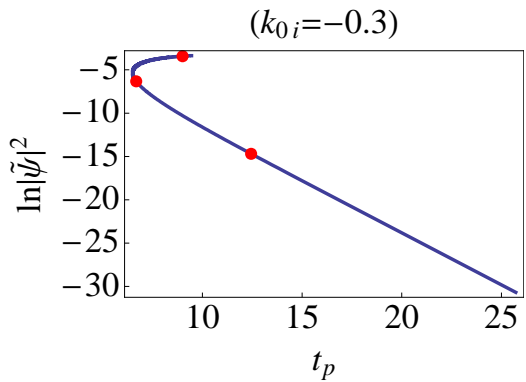}
 \end{center}
\caption{\label{cruce} 
Probability density versus $t_p(x)$ at the transition point for fixed $k_0=1-0.3i$. 
The drawing covers $x$ running from $0.0001$ to $13$. The dots correspond to the 
$x$ values considered in Fig. \ref{denx}.}
\end{figure}
%
Several interesting conclusions can be inferred from these figures
and the corresponding equations (\ref{eq42},\ref{eq43},\ref{eq45}). 
An important one is that the probability density at the transition point 
increases, as the observation point is moved away from the source, improving 
the possibility of observing the transition. 
A basic reason for this is the asymptotic growth of $|\psi_s|^2\sim x^2$, 
whereas the pole term density (straight line in Fig. \ref{denx}) 
is basically shifted by $\delta x/2k_{0R}$
for a shift $\delta x$ in the observation 
coordinate:\footnote{The onset of the exponential term is also delayed as determined by 
${{\rm{Im}}u_0^{(+)}=0}$.} in other words, the exponential behaviour is delayed by increasing 
$x$.\footnote{This is one of the main differences from the evanescent case
$\omega_{0R}<0$, for which the pole crossed by the steepest descent path is at 
$-k_0$, and the exponential term is {\it advanced} in time by increasing $x$.}

\begin{figure}[ht]
\begin{center}
\includegraphics[height=5cm,angle=0]{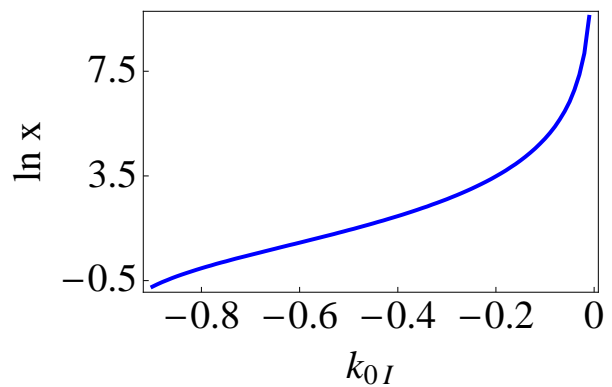}
\end{center}
\caption{\label{criticpos}
Largest $x$ value where an exponential to post-exponential transition is 
found, satisfying the conditions $R=1$ and ${\rm{Im}}u_0^{(+)}>0$,  versus $k_{0I}$ (normalized waves). 
}
\end{figure}

\begin{figure}[ht]
\begin{center}
\includegraphics[height=5cm,angle=0]{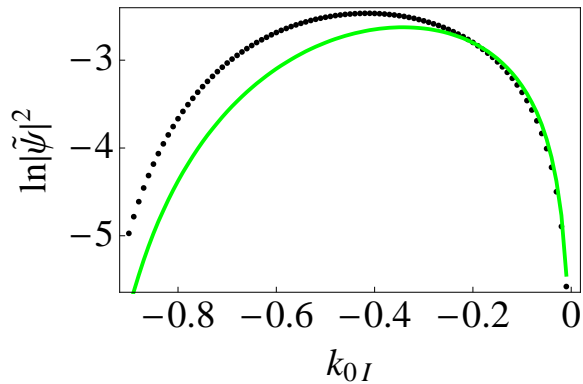}
\end{center}
\caption{\label{criticden}
(Color online) Density of normalized waves, at the largest $x$ value where the transition is defined,
as function of $k_{0I}$. Solid (green) line: density of the approximate solution; 
Dotted (black) line: value from the exact wavefunction.   
}
\end{figure}

Note also the behaviour of the transition time $t_p$ as $x$ increases. Initially 
$t_p$ occurs earlier as $x$ increases, but there is an $x$ value, $x=1/|k_{0I}|$,  
beyond which the tendency changes, and $t_p$ then increases with $x$ until the transition is no longer observable, 
as the onset of the exponential decay contribution is below the saddle contribution. 
Figure \ref{cruce} shows the density at the transition versus $t_p$. The curve depends parametrically 
on $x$, which increases from the bottom right corner upwards.
The last point marks the largest value of $x$ for which the transition may be defined by the $R=1$ criterion,
and such that ${\rm Im}(u_{0}^{(+)})>0$. This value of $x$ corresponds very nearly to 
the case depicted by dashed lines in Fig. \ref{denx}. 
This maximum $x$ increases with 
the resonance lifetime $\tau_0=1/|4k_{0I}|$, and it is plotted versus $|k_{0I}|$ in Fig. \ref{criticpos}. 
The (normalized, see Sec. \ref{normali}) density at that critical point is shown 
in Fig. \ref{criticden}. We can infer from this figure that there is a maximum 
of the transition point density slightly to the right of $k_{0I}\sim -0.5$. 
This is consistent with the condition described by Jittoh \cite{Jittoh}
for purely non-exponential decay,    
\beq						
 {\Gamma}/{\epsilon}=  {1}/[{k_{0R}^{2}\tau_0}] = 4|k_{0I}|\geq 2,
\eeq
where $\Gamma$ and $\epsilon$ are the (dimensionless) decay rate and real energy of the resonance, 
and we have used the fact that the real part $k_{0R}=1$ in our units.   

We are now better prepared to come back to Fig. \ref{3dim} and appreciate several of its 
features: note the exponential decay at the $x=0$ edge, the linear delay of the approximate 
wave front when $x$ is increased, the increase of the algebraic long-time tail with $x$, 
and the characteristic transition fringes at intermediate $x$ values, 
which disappear for larger $x$ values, characterized by purely non-exponential behavior.        
%
%
%
%
%
%
\section{Discussion}
The transition from exponential to non-exponential decay for an isolated decaying system 
has never been observed despite the well-established theoretical prediction that it must 
eventually occur. Apart from the difficulty of avoiding interactions that may extend the exponential region, 
a fundamental reason is the small value of the survival probability at the predicted transition.     
We have shown that observing the density of decay products at an optimal distance from the source, 
significantly improves the observability of this elusive transition. 
Almost 50 years ago R. G. Newton \cite{Newton} gave  a classical-trajectory argument 
pointing out that the transition would be affected by the observation distance. 
His predicted behaviour and dependence on parameters do not, however, coincide with 
the present quantum results. We have put forward a  simple decay model
that provides analytical expressions, facilitates predictions, and allows 
analysis of a proposed experiment through its dependence on resonance and 
observation parameters (release energy and lifetime, distance and time of density measurement). 
Experiments may be undertaken using 
quantum systems such as a leaking quantum dot, or escape of ultracold atoms from a trap. In 
both cases, trap and escape parameters are controllable, and it is possible to set a time zero 
for preparation of the decaying system. An effective 1D waveguide (a quantum wire for electrons, or a detuned laser for atoms)
may channel the escaping particle away from the trap. Localized, or spatially resolved particle
detection can be performed, e.g., with fluorescence or absorption techniques.    
Experimentally accessible low temperatures and conditions determine decoherence lengths 
and times, large enough to observe the effects described.  Atom-atom 
interactions may be minimized by using 
Feshbach resonances. As an example,  
let us consider an ${}^{87}$Rb quasi-one dimensional Bose-Einstein condensate sample of $10^6$ atoms, 
released to a quasi-1D waveguide from a controllable trap 
that will determine the resonance energy and lifetime \cite{prepa}.   
Then a commercial CCD camera with $\sim 3$ $\mu$m spatial resolution and shot noise limited detection 
with at least $\sim 10$ atoms/pixel may be used to measure the 
atomic density at the waveguide by resonant absorption imaging \cite{kruger}. 
According to our model a sample of $10^6$ atoms with lifetime $\sim 400$ $\mu$s, 
and release velocity $\sim 1$ cm/s would provide a $9200$ atoms/pixel transition density at 
$t_p\sim10$ ms and at a distance of $\sim 100$ $\mu$m. These numbers 
make the observation of the transition very plausible. Note that even $10^3$ initial atoms would be close to 
the shot-noise observation limit at that space-time point.  
An analysis of the effect of atom-atom interactions in the condensate 
is left for a separate study but, it is expected that 
such interactions will not be significant in the transition densities.    

A different physical realization may be possible through an array of long parallel waveguides 
\cite{Jo65, CLS03}. The variation with $z$, (the longitudinal 
distance along the guide) of the electric field in the $n$'th guide, 
plays the role of the variation with time of the site amplitudes
of a quantum tight-binding model \cite{Lon06a,VLL07}.
Recent experiments with scanning tunnelling microscopy have allowed Longhi and 
collaborators~\cite{VLL07} to measure the evanescent fields along the 
waveguides, and demonstrate  the classical-quantal analogy for a periodic parallel 
array. In addition~\cite{Bia08} the classical analogue of the quantum 
Zeno effect has been verified. 
  
If the coupling between adjacent sites, except for the first two, is described by 
a universal hopping constant, and the site energies are taken all to be equal, 
the Hamiltonian is, in dimensionless units, 
\begin{eqnarray}		     
H &=& -\Delta \bigg[ |1 \ra\la 2| + |2 \ra\la 1|\bigg]  \nonumber \\ 
&-&  \sum_{n=2}^{\infty}\, \bigg[ |n \ra\la n+1|+ |n+1 \ra\la n| \bigg], 
\label{eq:ini1}
\end{eqnarray}
where we take $0<\Delta <1$ to provide isolation of site $1$ from the rest of the array. 

When only site $1$ is occupied at time zero, the analytic solution for the amplitude at site $n$
shows a  
transition from exponential to power-law decay, at the time  
defined by equating the exponential contribution to the asymptotic power-law decay expression,   
dropping a sinusoidal oscillation peculiar to this model. 
\begin{eqnarray}			
t^{3/2}=\frac{[n+\alpha^2(n-2)]} {\sqrt{\pi}(1+\alpha^2)^3}{2 \alpha^{n+1}}e^{\gamma t/2}. 
\label{eq:enh3}
\end{eqnarray}
where $\alpha^2=1-\Delta^2$, and $\gamma=2\Delta^2/\alpha$. For small $\Delta$, $|\alpha| \sim 1$, making 
$\gamma << 1$, and  giving a long lifetime for exponential decay. But so long as $|\alpha| < 1$, 
$|\alpha|^{n+1} << 1$, for large enough $n$, allowing a solution to be found.  
(The oscillations in the power-law decay cause some imprecision in the transition time, similar to 
that discussed in relation to Fig. \ref{3dim}  for the  model of this paper.)
Eq. (\ref{eq:enh3}) is very similar to Eq. (\ref{eq45}), and predicts the same phenomena. 
This provides support for the general validity of our main conclusion, namely, that the 
optimal transition density increases with increasing distance of the detector from the trap. 

\section*{Acknowledgments}
We acknowledge discussions with A. del Campo and D. Steck, as well as funding by the
Basque Country University UPV-EHU (GIU07/40),  the 
Ministerio de Educaci\'on y Ciencia (FIS2006-10268-C03-01, and C03-02), 
and by NSERC-Canada (Discovery grant RGPIN-3198, DWLS). ET acknowledges financial support by the 
Basque Government (BFI08.151).


\begin{thebibliography}{10}
\bibitem{isr}J. G. Muga, F. Delgado, A. del Campo, and G. Garc\'\i a-Calder\'on, 
Phys. Rev. A {\bf 73}, 052112 (2006), and references therein. 

\bibitem{FGR78} L. Fonda, G.C. Ghirardi, and   A. Rimini, 
Rep. Prog. Phys. {\bf 41}, 587 (1978).
\bibitem{FP08} P. Facchi and S. Pascazio, J. Phys. A: Math. Gen {\bf 41}, 493001 (2008).
\bibitem{book} J. Martorell, J. G. Muga, and D. W. L. Sprung, in Time in Quantum Mechanics, Vol. 2, Springer, Berlin, 2009.
%
\bibitem{Nik68} N. N. Nikolaev, Sov. Phys. Usp. {\bf 11}, 522 (1968). 
\bibitem{No88} E. B. Norman, S. B. Gazes, S. G. Crane, and D. A. Bennett, 
Phys. Rev. Lett. {\bf 60}, 2246 (1988).  
\bibitem{Gr88} P. T. Greenland, Nature {\bf 335}, 298 (1988). 
\bibitem{Nghiep98} T. D. Nghiep, V. T. Hahn,  N. N. Son, 
Nucl. Phys. B (Proc. Suppl.) {\bf 66}, 533 (1998).
\bibitem{No95} E. B. Norman, B. Sur, K. T. Lesko, R. M. Larimer, D. J. DePaolo,   
and T. L. Owens, Phys. Lett. B {\bf 357}, 521 (1995). 
\bibitem{Greenland86} P. T. Greenland, Phys. Lett. A {\bf 117}, 181 (1986).
\bibitem{Law02b} R. E. Parrott and  J. Lawrence, Europhys. Lett. {\bf 57}, 632 (2002).
\bibitem{RHM06} C. Rothe, S. I. Hintschich,  and A. P. Monkman, Phys. 
Rev. Lett. {\bf 96}, 163601 (2006).
\bibitem{Winterh} R. G. Winter,  Phys. Rev. {\bf 126}, 1152 (1962).
\bibitem{KD08} L. M. Krauss and J. Dent, Phys. Rev. Lett. {\bf 100}, 171301
(2008).
\bibitem{DS84} S. D. Druger and  M. A. Samuel, Phys. Rev. A {\bf 30}, 640 (1984).
\bibitem{NB77} C. A. Nicolaides and  D. R. Beck, Phys. Rev. Lett. {\bf 38}, 683, 1037
(1977)
\bibitem{Nicolaides02} C. A. Nicolaides, Phys. Rev. A {\bf 66}, 022118 (2002).
\bibitem{MMS08} J. Martorell, J. G. Muga, and  D. W. L. Sprung, 
Phys. Rev. A {\bf 77}, 042719 (2008).
\bibitem{Lewenstein} M. Lewenstein and  K. Rzazewski, Phys. Rev. A {\bf 61}, 022105
(2000). 
\bibitem{Lon06a} S. Longhi, 
Phys. Rev. Lett. {\bf 97}, 110402 (2006).
\bibitem {VLL07} G. Della Valle, S. Longhi, P. Laporta, P. Biagioni,
L. Dou, and M. Finazzi, 
Appl. Phys. Lett. {\bf 90}, 261118 (2007).    
\bibitem{Jittoh} T. Jittoh, S. Matsumoto, J. Sato, Y. Sato,  K. Takeda,
Phys. Rev. A {\bf 71}, 012109 (2005).
\bibitem{GV06} G. Garc\'\i a-Calder\'on,  J. Villavicencio, 
Phys. Rev. A {\bf 73}, 062115 (2006). 
\bibitem{RLE82} K. Rzazewski, M. Lewenstein,  J. H. Eberly, J. Phys. B: At. Mol. Phys. {\bf 15}, L661 (1982).
\bibitem{ZRL84} J. Zakrzewski, K. Rzazewski,  M. Lewenstein, 
J. Phys. B: At. Mol. Phys. {\bf 17}, 729 (1984). 
\bibitem{Kelkar} N. G. Kelkar, M. Nowakowski, and K. P. Khemchandani, 
Phys. Rev. C {\bf 70}, 024601 (2004). 
\bibitem{DDGMR06} A. del Campo, F. Delgado, G. Garc\'\i a-Calder\'on, J. G. Muga 
and M. G. Raizen, Phys. Rev. A {\bf 74}, 013605 (2006).
\bibitem{Newton} R. G. Newton, Annals of Physics (NY) {\bf 14}, 333 (1961).
\bibitem{MB} J. G. Muga and  M. B{\"{u}}ttiker, Phys. Rev. A {\bf 62}, 023808 (2000). 
\bibitem{FMGG90} J. Felber, G. M\"{u}ller, R. G{\"{a}}hler, and R. Golub, Physica B {\bf 162}, 191 (1990).
\bibitem{BSCK92} H. R. Brown, J. Summhammer, R. E. Callaghan and P. Kaloyerou, Phys. Lett. A, {\bf 163}, 21 (1992).
\bibitem{BZ97} C. Brukner and A. Zeilinger, Phys. Rev. A {\bf 56}, 3804 (1997).
\bibitem{Stevens} K. W. H. Stevens, Eur. J. Phys. {\bf 1}, 98 (1980); 
J. Phys. C {\bf 16}, 3649 (1983).
\bibitem{Ranfa90} A. Ranfangi, D. Mugnai, P. Fabeni, and P. Pazzi, Phys. Scr. {\bf 42}, 508 (1990).
\bibitem{Ranfa91} A. Ranfangi, D. Mugnai, and A. Agresti, Phys. Lett. A {\bf 158}, 161 (1991).
\bibitem{Mor92} P. Moretti, Phys. Scripta {\bf 45}, 18 (1992).
\bibitem{BT98} M. B{\"{u}}ttiker, H. Thomas, Superlatt. Microstruct. {\bf 23}, 781 (1998).
\bibitem{DMRGV} F. Delgado, J. G. Muga, A. Ruschhaupt, G. Garc\'\i a-Calder\'on, J. Villavicencio, Phys. Rev. A {\bf 68} (2003) 032101.

\bibitem{DMR04} F. Delgado, J. G. Muga, A. Ruschhaupt, Phys. Rev. A {\bf 69}, 022106 (2004).

\bibitem{DM05} A. del Campo and  J. G. Muga, J. Phys. A, {\bf 38}, 9802 (2005).
\bibitem{DMM07} A. del Campo, J. G. Muga, and M. Moshinsky, J. Phys. B {\bf 40}, 975 (2007).
\bibitem{BEM01} A. D. Baute, I. L. Egusquiza, J. G. Muga, J. Phys. A {\bf 34}, 4289 (2001).
\bibitem{FT61} V. N. Faddeyeva, N. M. Terentev,
Mathematical Tables: Tables of the values of the function $w(z)$ for complex argument, Pergamon, New York, 1961.
\bibitem{AS65} A. Abramowitz, I. A. Stegun,
Handbook of Mathematical Functions, Dover, New York, 1965.
\bibitem{prepa} F. Delgado, J. G. Muga, and A. Ruschhaupt, Phys. Rev. A {\bf 74}, 063618 (2006). 
\bibitem{kruger} P. Kr\"uger, S. Wildermuth, S. Hofferberth, L. M. Andersson, S. Groth, I. Bar-Joseph, J. Schmiedmayer, J. Phys. Conference Series {\bf 19} (2005) 56-65.
\bibitem{Jo65} A. L. Jones, J. Opt. Soc. Am. {\bf 55}, 261-271 (1965).
\bibitem{CLS03} D. N. Christodoulides, F. Lederer and Y. Silverberg, 
Nature {\bf 424}, 817-23 (2003). 
\bibitem{Bia08} P. Biagoni, G. Della Valle, M. Ornigotti, M. Finazzi,
L. Duo, P. Laporta and S. Longhi, Optics Express {\bf 16}, 3762-7 (2008).
\end{thebibliography}
\end{document}